\documentclass[reqno, a4paper]{amsart}
\usepackage{amsmath}
\usepackage{amssymb}
\usepackage{amsthm}

\usepackage[scale=0.9]{geometry}

\usepackage{natbib}
\usepackage{bibentry} 

\usepackage[english]{babel}
\usepackage[utf8]{inputenc}

\usepackage{subfig}
\usepackage{graphicx}

\usepackage[unicode]{hyperref}
\usepackage[active]{srcltx} 

\usepackage{mathbbol}
\usepackage{bm} 

\usepackage{units}
\usepackage{tensor}
\usepackage{accents}

\usepackage{enumitem}

\newcommand{\exponential}[1]{\ensuremath{{\mathrm e}^{#1}}}

\newcommand{\iunit}{\ensuremath{\mathrm{i}}}

\newcommand{\bydefinition}{\mathrm{def}}

\newcommand{\diff}{\mathrm{d}}

\renewcommand{\vec}[1]{\ensuremath{\mathbf{#1}}}

\makeatletter
\@ifpackageloaded{bm}%
{\renewcommand{\vec}[1]{\ensuremath{\bm{#1}}}%
}{%
\relax
}
\makeatother
\makeatletter
\@ifpackageloaded{bm}%
{%
}{%

}

\@ifpackageloaded{bm}%
{%
 
}{%

}

\@ifpackageloaded{bm}%
{%
}{%

}
\makeatother


\newcommand{\Z}{\ensuremath{{\mathbb Z}}}

\newcommand{\pd}[2]{\ensuremath{\frac{\partial {#1}}{\partial {#2}}}}
\newcommand{\ppd}[2]{\ensuremath{\frac{\partial^2 {#1}}{\partial {#2^2}}}}
\newcommand{\dd}[2]{\ensuremath{\frac{\diff {#1}}{\diff {#2}}}}

\newcommand{\ddd}[2]{\ensuremath{\frac{\diff^2 {#1}}{\diff {#2}^2}}}

\makeatletter
\@ifpackageloaded{tensor}
{

}{%

}
\makeatother

\makeatletter
\@ifpackageloaded{tensor}
{

}{%

}
\makeatother

\makeatletter
\@ifundefined{volume}{%
}%
{%
}
\makeatother

\makeatletter

\@ifpackageloaded{MnSymbol} 
{
 
}{%
 
}

\@ifpackageloaded{MnSymbol} 
{
   
}{%
   
}
\makeatother

\newcommand{\meff}{m_{\text{eff}}}
\newcommand{\msum}{m_{\text{sum}}}

\newcommand{\mij}[2]{m_{#1}^{(#2)}}
\newcommand{\uij}[2]{u_{#1}^{(#2)}}
\newcommand{\sigmaj}[1]{\sigma^{(#1)}}

\newcommand{\Wj}[1]{W^{(#1)}}
\newcommand{\Tj}[1]{T^{(#1)}}

\numberwithin{equation}{section}

\title[Negative mass]{The conclusion that metamaterials could have negative mass is a consequence of improper constitutive characterisation}

\date{\today}

\author{D. Cichra}
\address{
Faculty of Mathematics and Physics\\
Charles University\\
Sokolovsk\'a 83\\
Praha 8 -- Karl\'{\i}n\\
CZ 186\;75\\
Czech Republic
}
\email{cichra.david@gmail.com}

\author{V. Pr\r{u}\v{s}a}
\address{
Faculty of Mathematics and Physics\\
Charles University\\
Sokolovsk\'a 83\\
Praha 8 -- Karl\'{\i}n\\
CZ 186\;75\\
Czech Republic
}
\email{prusv@karlin.mff.cuni.cz}

\author{K. R. Rajagopal}
\address{
Texas A\&M University\\
Department of Mechanical Engineering\\
3123 TAMU\\
College Station TX 77843-3123\\
United States of America
}
\email{krajagopal@tamu.edu}

\author{C. Rodriguez}
\address{
University of North Carolina\\
Department of Mathematics\\
329 Phillips Hall \\
Chapel Hill NC 27599 \\
United States of America
}
\email{crodrig@email.unc.edu}

\author{M. Vejvoda}
\address{
Faculty of Mathematics and Physics\\
Charles University\\
Sokolovsk\'a 83\\
Praha 8 -- Karl\'{\i}n\\
CZ 186\;75\\
Czech Republic
}
\email{martin.vejvoda@mff.cuni.cz}

\thanks{V. Pr\r{u}\v{s}a thanks the Czech Science Foundation, grant no. 20-11027X, for its support. C. Rodriguez and K. R. Rajagopal thank the National Science Foundation, grant DMS-2307562, for its support.}

\begin{document}

\begin{abstract}
  The concept of ``effective mass'' is frequently used for the simplification of complex lumped parameter systems (discrete dynamical systems) as well as materials that have complicated microstructural features. From the perspective of wave propagation, it is claimed that for some bodies described as metamaterials, the corresponding ``effective mass'' can be frequency dependent, negative or it may not even be a scalar quantity. The procedure has even led some authors to suggest that Newton’s second law needs to be modified within the context of classical continuum mechanics. Such absurd physical conclusions are a consequence of appealing to the notion of ``effective mass'' with a preconception for the constitutive structure of the metamaterial and using a correct mathematical procedure. We show that such unreasonable physical conclusions would not arise if we were to use the appropriate ``effective constitutive relation'' for the metamaterial, rather than use the concept of ``effective mass'' with an incorrect predetermined constitutive relation.



\end{abstract}

\maketitle

\tableofcontents


\section{Introduction}
\label{sec:introduction}
According to the Oxford English Dictionary\footnote{Oxford English Dictionary, s.v. “metamaterial (n.),” July 2023, https://doi.org/10.1093/OED/2009897523}, the terminology ``metamaterials'' first occurs in the writings of J. Toribio in 1996, in his article titled ``Teaching fracture mechanics in civil engineering education: the Spanish experience''. In the dictionary we find the following definition to the terminology metamaterials: ``A synthetic composite material engineered to display properties not usually found in natural materials.'' In his paper \cite{toribio.j:teaching} delineates his 150 hour long teaching program for fracture mechanics and material science consisting in a variety of topics. He then states the program ``---results in a general evolution of knowledge from the simply-connected continuum media (continuum mechanics), including the particular theories of elasticity and plasticity, through the multi-connected continuum media (macro-aspects of fracture), the discontinuous or disconnected media (micro-aspects of fracture and material science), and finally materials made with other materials, which can be named meta-materials (composite materials).'' The tacit assumption that Toribio is making is that the composite material under consideration can be genuinely described as a continuum, with meaningfully defined material properties, so that one can study the problem of fracture and fatigue of such materials.

A spate of publications concerning ``metamaterials'' have appeared recently as they supposedly exhibit astonishing, and some of them, bizarre properties, see, for example, \cite{milton.gw.willis.jr:on}. These conclusions concerning their behavior are more a consequence of mistaken and unsound interpretation of their response characteristics stemming from faulty constitutive modeling of such materials, than metamaterials truly displaying such attributes. Given a body, a modeler, based on how the body responds to a variety of inputs, decides on the appropriate constitutive relation to describe its behavior and studies its response to arbitrary inputs, subject to accepted laws of physics, which in the case of continua within the context of classical Newtonian mechanics are the balance of mass, linear and angular momentum, energy and the second law of thermodynamics. (We are making the tacit assumption that we are not concerned with magnetic, electric and other fields and hence ignoring  other field equations.) If the assumed constitutive relation does not comply with accepted physical principles, then it would be prudent to look for a more appropriate constitutive relation to describe the response of the body, instead of insisting in the use of an erroneous constitutive relation and propose that the principles of physics, which has served us well, need to be rethought. While it is true that when dealing with atomic or sub-atomic particles, and other such areas, we do have to go outside the ambit of classical Newtonian physics, this is definitely not warranted in the case of metamaterials that supposedly display such unacceptable physical characteristics. Our study shows that we can develop constitutive relations that can describe the response of metamaterials which do not exhibit the kind of unreasonable physical characteristics as those presented in the recent literature concerning metamaterials, namely the~\emph{negative effective mass}. 

The metamaterials that have been shown to display unreasonable physical characteristics---\emph{negative effective mass}---are essentially lumped parameter systems, with continua being comprised of homogenization of such systems. Lumped parameter systems have been used to model continua as they simplify the governing equations of the body under consideration, which are usually partial differential equations or integro-partial differential systems, to the study of ordinary differential equations, thereby making the problem tractable and amenable to analysis. For instance, certain elastic bodies are described as a lumped parameter system consisting in springs and masses, while viscoelastic fluid and solid bodies have been described by lumped parameter systems comprising of springs, dashpots and masses, with inelastic solids being considered lumped parameter systems with a dissipative mechanism that is appropriate for sliding solids. While continua have been approximated using lumped parameter systems, the converse, that is, how a homogenized lumped parameter system might be appropriately described as a continuum with meaningful physical properties, is a rather subtle issue. Not exercising due care could lead one to absurd conclusions such as the effective mass of the continuum that replaces the lumped parameter system being negative or it not being a scalar quantity, as has been observed in the case of metamaterials, see~\cite{milton.gw.willis.jr:on} to name a few. This paper is devoted to elucidating how one determines an \emph{effective constitutive relation} to describe the lumped parameter system that avoids pitfalls such as the \emph{effective mass} of the body being negative.

The concept of \emph{effective quantities} allows one to model the behaviour of complex systems as if they were simpler ones. Examples thereof known to every physics student are the period of oscillations formula for a \emph{physical pendulum} (a rigid rod that undergoes fixed axis rotation about a fixed point) interpreted as a period of oscillations formula for the \emph{mathematical pendulum} (a point mass attached to a massless rod rotating about a fixed point) wherein the rod \emph{length} for the mathematical pendulum is modified accordingly, see, for example, \cite[Section 14.6]{young.hd.freedman.ra.ea:university}. Another example is the \emph{effective mass} used in the formula for the period of oscillations for a point mass attached to a \emph{massless spring} versus the period of oscillations for the point mass attached to a \emph{spring with a given linear density}, see, for example, \cite[Exercise 14.93]{young.hd.freedman.ra.ea:university} and in-depth discussion in~\cite{christensen.j:improved}. While the use of effective quantities is fine in the aforementioned cases, an absurd conclusion concerning the concept of \emph{effective mass} appeared within the context of metamaterials.

Some metamaterials, that is materials with complicated engineered microstructure, are described using the \emph{effective mass concept}, see, for example, \cite{chan.ct.li.j.ea:on}, \cite{li.j.chan.ct:double-negative}, \cite{liu.z.chan.ct.ea:analytic}, \cite{sheng.p.zhang.xx.ea:locally}, \cite{milton.gw.willis.jr:on} and \cite{huang.hh.sun.ct.ea:on}. In these cases the effective mass concept is used to study especially the elastic wave transmission, and unlike in the standard cases discussed above the effective mass/density is for some metamaterials \emph{non-constant}. The effective mass can depend on the forcing frequency, and it can be, for certain forcing frequencies, even~\emph{negative}. This is a startling observation, and it lead some authors to even claim that this observation requires one to modify Newton's laws. While such a situation is warranted when dealing with the dynamics of atomic and sub-atomic particles, it is not justified in  studies such as those carried out by~\cite{milton.gw.willis.jr:on} as they are cast within the purview of Newtonian mechanics. For example, \cite{milton.gw.willis.jr:on} claim that

\begin{quote}
  In our models, the motion of the rigid material apparently violates Newton’s law owing to the vibration of the internal masses: ‘force equals mass times acceleration’ only applies if mass is replaced by ‘effective mass’, which is a non-local operator in time (equivalently, a function of frequency under any purely harmonic forcing). Furthermore, unless the microstructure is specially constructed so as to be isotropic, the ‘effective mass’ is a second-order tensor, not a scalar.
\end{quote}

We emphasise the analysis in \cite{milton.gw.willis.jr:on} and similar papers such as \cite{huang.hh.sun.ct.ea:on} is---from the perspective of routine mathematical manipulations---correct. However, as \cite{schwartz:pernicious} aptly observes:

\begin{quote}
Related to this deficiency of mathematics, and perhaps more productive [of] rueful consequence, is the simple-mindedness of mathematics—its willingness, like that of a computing machine, to elaborate upon any idea, however absurd; to dress scientific brilliancies and scientific absurdities alike in the impressive uniform of formulae and theorems. [\ldots] The result, perhaps most common in the social sciences, is bad theory with a mathematical passport.
\end{quote}

The routine mathematical manipulations indeed suggests that one must introduce the frequency dependent and possibly negative \emph{effective mass}, and that the concept of Newton's laws and classical continuum mechanics must be questioned. In the present contribution we however show that there exists a \emph{different interpretation of the mathematical manipulations} involved in the analysis of metamaterials, and that this interpretation is in conformity with Newton's laws and the classical concept of mass/density as a scalar, positive and constant quantity.

The key idea is a simple one. If Newton's second law reads
\begin{equation}
  \label{eq:1}
  m \ddd{\vec{x}}{t} = \vec{F},
\end{equation}
and if we want to keep the mass a positive constant scalar quantity, then the only other possibility is to modify the force. The fact that there exist different expressions for the force is commonly accepted. For example, the spring can be a linear spring or cubic spring, the response of a solid in the nonlinear setting (finite deformations) can be given by the neo-Hookean constitutive relation or by a more complicated constitutive relation, see, for example, \cite{rivlin.rs:large*1} and \cite{truesdell.c.noll.w:non-linear}, and recent lists of various constitutive relations in~\cite{mihai.la.goriely.a:how} or \cite{destrade.m.saccomandi.g.ea:methodical} to name a few. We choose to follow this line of reasoning.

If the mass must be kept intact, we have to work with an effective constitutive relation, that is with stress/force versus strain/displacement relation. The idea is in fact the same as in the theory of viscoleastic fluids and solids, see, for example~\cite{bland.dr:theory}, \cite{wineman.as.rajagopal.kr:mechanical}, or \cite{wineman.a:nonlinear}, wherein the behaviour a complex mechanical analog made of springs and dashpots is described by a single effective constitutive relation. (This can be a complicated algebraic, rate-type or integral equation.) This effective constitutive relation then works only with the stress/force and the strain/displacement and the derivatives of these quantities, and it completely hides the internal arrangement of the individual springs and dashpots. We want to achieve the same in the case of metamaterials, that is we want to identify the \emph{effective constitutive relation} that would allow us to describe the behaviour of metamaterials without referring to the concept of \emph{effective negative mass}. 


\section{Model systems}
\label{sec:model-system}

\cite{huang.hh.sun.ct.ea:on} introduce two simple model mechanical systems that document the concept of negative mass, namely they deal with a \emph{single mass-in-mass system} and \emph{one dimensional mass-in-mass lattice system}. We first recall the standard reasoning that leads to the negative mass concept---here we closely follow the presentation in~\cite{huang.hh.sun.ct.ea:on}---and then we proceed with the alternative interpretation based on the concept of an effective constitutive relation.

\subsection{Single mass-in-mass system}
\label{sec:single-mass-mass}
\begin{figure}[h]
  \centering
  \subfloat[\label{fig:mass-in-mass-a}Fully resolved system with visible inner mass $m_2$.]{\includegraphics[width=0.3\textwidth]{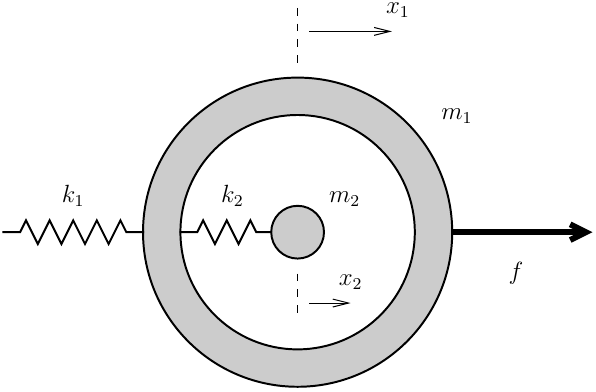}}
  \qquad
  \subfloat[\label{fig:mass-in-mass-b}Effective single mass system, effective mass $\meff$.]{\includegraphics[width=0.3\textwidth]{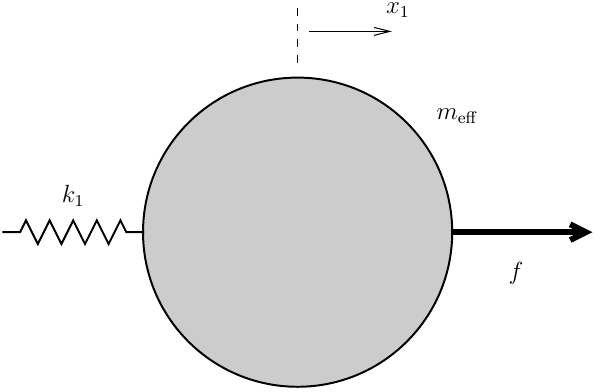}}
  \qquad
  \subfloat[\label{fig:mass-in-mass-c}Effective single mass system, effective linear constitutive relation $\sigma = \mathcal{L}(x_1)$.]{\includegraphics[width=0.3\textwidth]{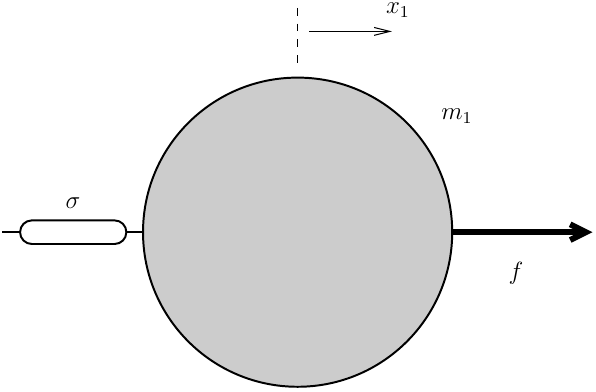}}
  \caption{Mass-in-mass model.}
  \label{fig:mass-in-mass}
\end{figure}

The standard model system is the \emph{mass-in-mass} system shown in Figure~\ref{fig:mass-in-mass-a}. In this case the evolution equations read
\begin{subequations}
  \label{eq:2}
  \begin{align}
    \label{eq:3}
    m_1 \ddd{x_1}{t}  + k_1 x_1 + k_2 \left(x_1 - x_2\right) & = f, \\
    \label{eq:4}
    m_2 \ddd{x_2}{t}  + k_2 x_2 - k_2 x_1 & = 0,
  \end{align}
\end{subequations}
where $f$ is a given right-hand side, and $k_1$ and $k_2$ are constant\footnote{The force $f$ must be acting on the outer mass only, gravitational force is not included since it would be acting directly on the inner mass as well.}, see~\cite{huang.hh.sun.ct.ea:on}. (The system is the same as the ``undamped vibration absorber'' system analysed in~\cite[Section 5.9]{meirovitch.l:fundamentals}. This standard text on mechanical vibrations however does not use the concept of negative mass, the system is analysed as is without the ambition to develop a reduced model that hides the inner mass $m_2$.) Typically, one is mainly interested in the system response to periodic forcing,
\begin{equation}
  \label{eq:5}
  f = \widehat{F} \sin \left( \Omega t\right),
\end{equation}
and transient system response is ignored. For later use we introduce the notation
\begin{subequations}
  \label{eq:6}
  \begin{align}
    \label{eq:7}
    \omega_1 &=_{\bydefinition} \sqrt{\frac{k_1}{m_1}}, \\
    \label{eq:8}
    \omega_2 &=_{\bydefinition} \sqrt{\frac{k_2}{m_2}},
  \end{align}
\end{subequations}
for the natural frequencies. The objective is to write~\eqref{eq:2} as a single ordinary differential equation for the first body (outer mass) only, and this equation should still have the standard Newtonian form
\begin{equation}
  \label{eq:10}
  m \ddd{x}{t}  + k x = f
\end{equation}
or, the even more general form,
\begin{equation}
  \label{eq:11}
  m \ddd{x}{t}  - \sigma = f, 
\end{equation}
where the stress/force $\sigma$ is given by a constitutive relation for the corresponding material. Clearly, \eqref{eq:10} is the special case of~\eqref{eq:11} in the case of a simple linear Hookean spring.

\subsubsection{Standard approach}
\label{sec:standard-approach}
The standard argument goes as follows. If the right-hand side $f$ has the form~\eqref{eq:5}, then we can search for the solution in the form of
\begin{equation}
  \label{eq:12}
  x_i = \widehat{x_i}\sin \left( \Omega t \right),
\end{equation}
which is essentially the Fourier transform of the equations. If we accept the \emph{ansatz} and substitute into the governing equations~\eqref{eq:2}, then we get the system of equations
\begin{subequations}
  \label{eq:13}
  \begin{align}
    \label{eq:14}
    - m_1 \Omega^2 \widehat{x_1} + k_1  \widehat{x_1} + k_2 \widehat{x_1} - k_2 \widehat{x_2} &= \widehat{F}, \\
    \label{eq:15}
    - m_2 \Omega^2 \widehat{x_2}  + k_2 \widehat{x_2} - k_2  \widehat{x_1} & = 0.
  \end{align}
\end{subequations}
We solve the second equation~\eqref{eq:15} for $\widehat{x_2}$, and we get
\begin{equation}
  \label{eq:16}
  \widehat{x_2}
  =
  \frac{k_2}{k_2 - m_2 \Omega^2} \widehat{x_1},
\end{equation}
which upon substitution into~\eqref{eq:14} yields
\begin{equation}
  \label{eq:17}
  - m_1 \Omega^2 \widehat{x_1} + k_1  \widehat{x_1} + \left(k_2  - \frac{k_2^2}{k_2 - m_2 \Omega^2} \right) \widehat{x_1}  = \widehat{F}.
\end{equation}
A simple algebraic manipulation reveals that the last equation can be further rewritten as
\begin{equation}
  \label{eq:19}
  -
  \left(
    m_1 + m_2 \frac{\omega_2^2}{\omega_2^2 - \Omega^2} 
  \right)
  \Omega^2
  \widehat{x_1}
  +
  k_1  \widehat{x_1}
  =
  \widehat{F}
  ,
\end{equation}
where we have used the notation introduced in~\eqref{eq:6}. The equation \eqref{eq:19}---which is formulated in the \emph{frequency space} (Fourier space)---resembles the \emph{frequency space} (Fourier space) representation of the equation
\begin{equation}
  \label{eq:20}
  m \ddd{x}{t}  + k x = f,
\end{equation}
which reads
\begin{equation}
  \label{eq:21}
  - m \Omega^2 \widehat{x} + k \widehat{x} = \widehat{F}.
\end{equation}
The visual comparison of~\eqref{eq:21} and~\eqref{eq:19} then motivates one to introduce the \emph{effective mass}, denoted as $\meff$, as
\begin{equation}
  \label{eq:22}
  \meff
  =_{\bydefinition}
  m_1 + m_2 \frac{\omega_2^2}{\omega_2^2 - \Omega^2},
\end{equation}
see~\cite{huang.hh.sun.ct.ea:on}. One then writes~\eqref{eq:19} as
\begin{equation}
  \label{eq:23}
  -
  \meff
  \Omega^2
  \widehat{x_1}
  +
  k_1  \widehat{x_1}
  =
  F
  .
\end{equation}

At this juncture, it is relevant to make a few pertinent observations. First, we note that~\eqref{eq:23} is \emph{not} a frequency space representation of
\begin{equation}
  \label{eq:24}
  -
  \meff
  \ddd{x_1}{t}
  +
  k_1 x_1
  =
  f
  ,
\end{equation}
because the effective mass $\meff$ in~\eqref{eq:23} is \emph{frequency dependent}. Second, if the forcing frequency $\Omega$ is slightly above the natural frequency $\omega_2$, then the effective mass is \emph{negative}. This second property is absurd and motivates finding an alternative effective system for the outer mass that is equivalent to \eqref{eq:2}.

We reiterate that from the mathematical point of view, there is nothing wrong with the calculations starting from \eqref{eq:2} and concluding with \eqref{eq:19}; only the interpretation of the term on the left-hand side of equation~\eqref{eq:23} as the effective mass, and interpreting~\eqref{eq:24} as the equation of motion with the effective mass, is bizarre. We now show that the effective system for the outer mass that is equivalent to \eqref{eq:2} can still have the standard Newtonian law form~\eqref{eq:10} with \emph{mass} being a constant positive quantity.

\subsubsection{Approach based on rate-type constitutive relation}
\label{sec:approach-based-rate}
The mass and the external force are the quantities that are immutable, hence the only quantity we are allowed to change in the equation
\begin{equation}
  \label{eq:25}
  m \ddd{x}{t}  + k x = f
\end{equation}
is the constitutive relation for the spring. Instead of the linear Hookean spring
\begin{equation}
  \label{eq:26}
  \sigma = - k x,
\end{equation}
we use a generic stress/force $\sigma$ such that
\begin{equation}
  \label{eq:27}
  m \ddd{x}{t}  - \sigma = f.
\end{equation}
The task is to identify the constitutive relation for the stress/force $\sigma$ in such a way that the equation
\begin{equation}
  \label{eq:28}
  m_1 \ddd{x_1}{t}  - \sigma = f
\end{equation}
gives the same solution as the system~\eqref{eq:2}. This is easy to do, we basically employ a variant of the approach well known in theory of rate-type materials (viscoelastic fluids and viscoelastic solids), see, for example, \cite{wineman.as.rajagopal.kr:mechanical}. 

We define
\begin{equation}
  \label{eq:29}
  \sigma = _{\bydefinition}  -  k_1 x_1 - k_2 \left(x_1 - x_2\right),
\end{equation}
where, as before, $k_1$ and $k_2$ are constant. We observe that
\begin{equation}
  \label{eq:30}
  x_2 = \frac{1}{k_2} \left( \sigma + \left(k_1 + k_2\right) x_1 \right).
\end{equation}
If we differentiate the stress/force $\sigma$ defined by~\eqref{eq:29} with respect to time, then we get
\begin{subequations}
  \label{eq:31}
  \begin{align}
    \label{eq:32}
    \dd{\sigma}{t} &= -\left(k_1 + k_2\right) \dd{x_1}{t} + k_2 \dd{x_2}{t}, \\
    \label{eq:33}
    \ddd{\sigma}{t} &= -\left(k_1 + k_2\right) \ddd{x_1}{t} + k_2 \ddd{x_2}{t}.
  \end{align}
\end{subequations}
Making use of the evolution equation for the inner mass $m_2$, see~\eqref{eq:4}, we get
\begin{equation}
  \label{eq:34}
  \ddd{x_2}{t} = \frac{k_2}{m_2} \left(x_1 - x_2\right),
\end{equation}
which in virtue of~\eqref{eq:30} yields
\begin{equation}
  \label{eq:36}
  k_2\ddd{x_2}{t} = - \frac{k_2}{m_2} \sigma - \frac{k_1 k_2}{m_2} x_1.
\end{equation}
Now we substitute~\eqref{eq:36} into~\eqref{eq:33} and get a differential equation in terms of unknown functions $\sigma$ and~$x_1$,
\begin{equation}
  \label{eq:38}
  \ddd{\sigma}{t} + \frac{k_2}{m_2} \sigma = -\left(k_1 + k_2\right) \ddd{x_1}{t} - \frac{k_1 k_2}{m_2} x_1.
\end{equation}
This is a rate-type constitutive relation that gives us the stress/force $\sigma$ in terms of the displacement/strain $x_1$.

The previous calculations show that the original mass-in-mass system,
\begin{subequations}
  \label{eq:39}
  \begin{align}
    \label{eq:40}
    m_1 \ddd{x_1}{t}  + k_1 x_1 + k_2 \left(x_1 - x_2\right) & = f, \\
    \label{eq:41}
    m_2 \ddd{x_2}{t}  + k_2 x_2 - k_2 x_1 & = 0,
  \end{align}
\end{subequations}
is equivalent to the effective system
\begin{subequations}
  \label{eq:42}
  \begin{align}
    \label{eq:43}
    m_1 \ddd{x_1}{t}  - \sigma &= f, \\
    \label{eq:44}
    \ddd{\sigma}{t} + \frac{k_2}{m_2} \sigma &= -\left(k_1 + k_2\right) \ddd{x_1}{t} - \frac{k_1 k_2}{m_2} x_1.
  \end{align}
\end{subequations}
The insight is that the spring in the ``reduced'' system~\eqref{eq:42} does not behave as a linear Hookean spring; the constitutive relation between the stress/force $\sigma$ and the displacement/strain $x_1$ is a linear rate-type equation. We note the rate-type constitutive relations are well-established concept in the theory of constitutive relation in continuum mechanics, see~\cite{thomson.w:on}, \cite{maxwell.jc:on}, \cite{burgers.jm:mechanical} or \cite{oldroyd.jg:on} for early examples thereof.

We also observe that we can solve~\eqref{eq:44} explicitly via the variation of constants method and write
\begin{equation}
  \label{eq:45}
  \sigma = \mathcal{L}(x_1),
\end{equation}
where $\mathcal{L}$ is a linear operator given by the formula
\begin{subequations}
  \label{eq:46}   
  \begin{equation}
    \label{eq:47}
    \mathcal{L}(x_1)
    =
        \sigma_0 \cos \left(\omega_2 t\right) +
    \varsigma_0 \sin \left( \omega_2t \right)
    +
    \frac{1}{\omega_2} \int_{\tau = 0}^t \sin \left( \omega_2 (t-\tau) \right) g(\tau )\, \diff \tau
    ,
  \end{equation}
  where
  \begin{equation}
    \label{eq:48}
    g(\tau) =_{\bydefinition} -\left(k_1 + k_2\right) \left. \ddd{x_1}{t} \right|_{t = \tau} - \frac{k_1 k_2}{m_2} \left. x_1 \right|_{t = \tau}.
  \end{equation}
\end{subequations}
The initial conditions $\sigma_0 = \sigma |_{t = 0}$ and $\varsigma_0 = \dd{\sigma}{t} |_{t = 0}$ are determined from the initial conditions for $x_1$ and $x_2$ via \eqref{eq:29} and~\eqref{eq:32}. 
Once~\eqref{eq:45} is substituted into~\eqref{eq:43}, we get a reduced model in terms of a single equation
\begin{equation}
  \label{eq:49}
  m_1 \ddd{x_1}{t}  - \mathcal{L}(x_1) = f,
\end{equation}
with the stress/force $\sigma$ given by a linear operator acting on the displacement/strain. This situation is common in viscoelasticity wherein the force/stress is given either in terms of a rate-type evolution equation or in terms of an convolution integral, see, for example, \cite{wineman.as.rajagopal.kr:mechanical}.

The response of system~\eqref{eq:42} to the periodic forcing~\eqref{eq:5} is obtained via the standard Fourier transform technique. The governing equations~\eqref{eq:39} and~\eqref{eq:42} are equivalent, and hence, we must get the same results as the standard approach based on the negative mass concept. We present the analysis just for the sake of completeness. Assuming that $x_1 = \widehat{x_1}\sin \left( \Omega t \right)$ and $\sigma = \widehat{\sigma}\sin \left( \Omega t \right)$, we see that~\eqref{eq:42} simplifies to
\begin{subequations}
  \label{eq:50}
  \begin{align}
    \label{eq:51}
    - m_1 \Omega^2 \widehat{x_1}  - \widehat{\sigma} &= \widehat{F}, \\
    \label{eq:52}
    - \Omega^2 \widehat{\sigma} + \frac{k_2}{m_2} \widehat{\sigma} &= \left(k_1 + k_2\right) \Omega^2 \widehat{x_1} - \frac{k_1 k_2}{m_2} \widehat{x_1}.
  \end{align}
\end{subequations}
The second equation can be solved for $\widehat{\sigma}$, which yields
\begin{equation}
  \label{eq:53}
  \widehat{\sigma}
  =
  \frac{\left(k_1 + k_2\right) \Omega^2 - k_1 \omega_2^2}{\omega_2^2 - \Omega^2}
  \widehat{x_1},
\end{equation}
where we have used the notation introduced in~\eqref{eq:6}. Substituting~\eqref{eq:53} into~\eqref{eq:51} then gives
\begin{equation}
  \label{eq:54}
  -
  \left(
    m_1
    +
    m_2
    \frac{\omega_2^2}{\omega_2^2 - \Omega^2}
  \right)
  \Omega^2 \widehat{x_1}
  +
  k_1 \widehat{x_1} = \widehat{F}.
\end{equation}
This result is the same equation as~\eqref{eq:19}.

We note that since the system of equations~\eqref{eq:42} is equivalent to the system~\eqref{eq:39}, both systems share the same energy. The energy~$E$---that is the kinetic energy plus the elastic stored energy---for the system~\eqref{eq:39} reads
\begin{equation}
  \label{eq:55}
  E
  =_{\bydefinition}
  \frac{1}{2}
  m_1
  \left(
    \dd{x_1}{t}
  \right)^2
  +
  \frac{1}{2}
  m_2
  \left(
    \dd{x_2}{t}
  \right)^2
  +
  k_1
  \frac{x_1^2}{2}
  +
  k_2 \frac{\left(x_1 - x_2\right)^2}{2},
\end{equation}
and it is straightforward to see that the energy is constant along the solution of system~\eqref{eq:39} in the absence of external forcing ($f = 0$). We can make use of~\eqref{eq:32} and~\eqref{eq:29} and rewrite the energy in terms of the stress/force, displacement/strain and the first time derivatives of these quantities,
\begin{equation}
  \label{eq:56}
  E
  =
  \frac{1}{2}
  m_1
  \left(
    \dd{x_1}{t}
  \right)^2
  +
  \frac{1}{2}
  \frac{m_2}{k_2^2}
  \left(
    \dd{\sigma}{t} + \left(k_1 + k_2\right) \dd{x_1}{t}
  \right)^2
  +
  k_1
  \frac{x_1^2}{2}
  +
  \frac{\left( \sigma +  k_1 x_1\right)^2}{2k_2}.
\end{equation}
If $f = 0$, then $E$ is the conserved quantity for the system~\eqref{eq:42}.

Next we note that the equivalence between the effective system~\eqref{eq:39} and the original mass-in-mass system~\eqref{eq:42} holds in the physical space and even for initial value problems. There is no need to restrict oneself to the frequency space and periodic solutions. The initial conditions for~\eqref{eq:39} are
\begin{subequations}
  \label{eq:57}
  \begin{align}
    \label{eq:58}
    \left. x_1 \right|_{t=0} &= x_{1,0}, \\
    \label{eq:59}
    \left. \dd{x_1}{t} \right|_{t=0} &= v_{1,0}, \\
    \label{eq:60}
    \left. x_2 \right|_{t=0} &= x_{2,0}, \\
    \label{eq:61}
    \left. \dd{x_2}{t} \right|_{t=0} &= v_{2,0},
  \end{align}
\end{subequations}
while the initial conditions for~\eqref{eq:42} are
\begin{subequations}
  \label{eq:62}
  \begin{align}
    \label{eq:63}
    \left. x_1 \right|_{t=0} &= x_{1,0}, \\
    \label{eq:64}
    \left. \dd{x_1}{t} \right|_{t=0} &= v_{1,0}, \\
    \label{eq:65}
    \left. \sigma \right|_{t=0} &= \sigma_{0}, \\
    \label{eq:66}
    \left. \dd{\sigma}{t} \right|_{t=0} &= \varsigma_{0}.    
  \end{align}
\end{subequations}
The initial conditions for the stress/force $\sigma$---that is~\eqref{eq:65} and \eqref{eq:66}---are easy to obtain from~\eqref{eq:57} in virtue of relations~\eqref{eq:29} and~\eqref{eq:32}.

\subsection{One dimensional mass-in-mass lattice system}
\label{sec:lattice-mass-mass}

\begin{figure}[h]
  \centering
  \includegraphics[width=0.8\textwidth]{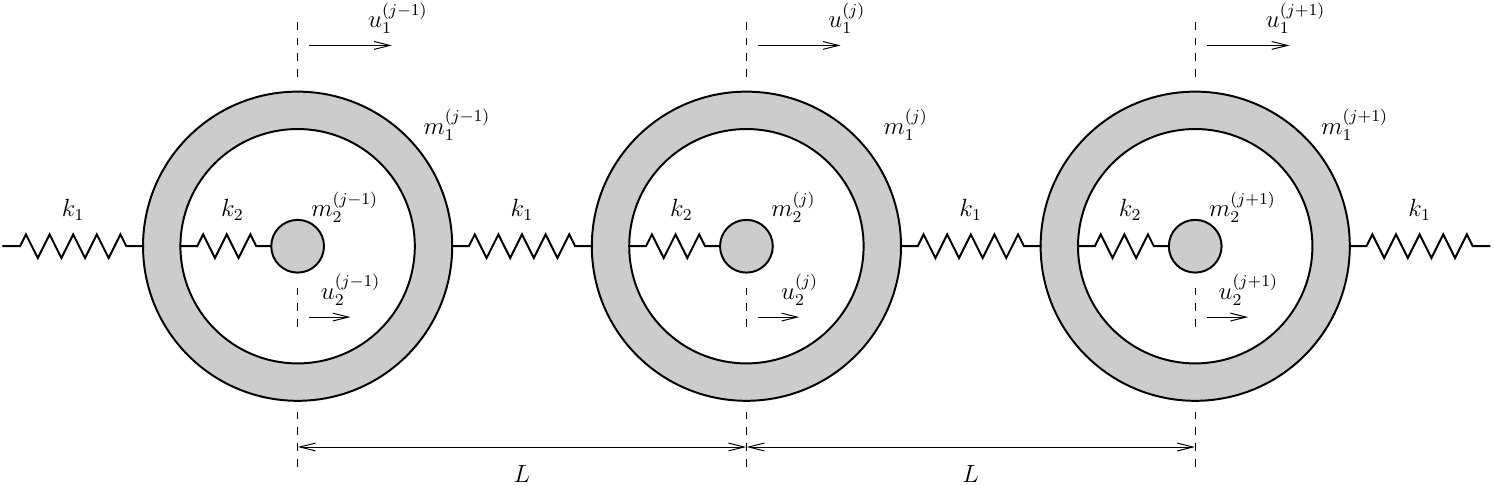}
  \caption{Infinite mass-in-mass one-dimensional lattice structure.}
  \label{fig:infinite-mass-in-mass}
\end{figure}
The next system studied by~\cite{huang.hh.sun.ct.ea:on} is the infinitely long one-dimensional lattice system consisting of single mass-in-mass systems, see Figure~\ref{fig:infinite-mass-in-mass}. The importance of this model system dwells in the fact that it serves as a bridge between discrete mass systems and systems with continuously distributed matter. Furthermore the one-dimensional lattice system allows one to study wave transmission phenomena. The governing equations for the $j$-th unit cell read
\begin{subequations}
  \label{eq:67}
  \begin{align}
    \label{eq:68}
    \mij{1}{j} \ddd{\uij{1}{j}}{t}  - k_1 \left(\uij{1}{j+1} - \uij{1}{j} \right) + k_1 \left(\uij{1}{j} - \uij{1}{j-1} \right) + k_2 \left(\uij{1}{j} - \uij{2}{j}\right) & = 0, \\
    \label{eq:69}
    \mij{2}{j} \ddd{\uij{2}{j}}{t}  + k_2 \left(\uij{2}{j} -  \uij{1}{j}\right) & = 0,
  \end{align}
\end{subequations}
see~\cite{huang.hh.sun.ct.ea:on}. We note that~\eqref{eq:68} can be rewritten as
\begin{equation}
	\label{eq:70}
	\mij{1}{j} \ddd{\uij{1}{j}}{t}  - k_1 \left(\uij{1}{j+1} - 2\uij{1}{j} +  \uij{1}{j-1} \right) + k_2 \left(\uij{1}{j} - \uij{2}{j}\right)  = 0.
\end{equation} The meaning of the individual terms in \eqref{eq:68} is clear: the term $k_1 \left(\uij{1}{j+1} - \uij{1}{j} \right)$ represents the interaction of $j$-th cell with its right neighbour, that is with the $(j+1)$-th unit cell, while  the term $k_1 \left(\uij{1}{j} - \uij{1}{j-1} \right)$ represents the interaction with its left neighbour, that is with the $(j-1)$-th unit cell. The term $ k_2 \left(\uij{1}{j} - \uij{2}{j}\right)$ describes the self-interaction with the inner mass. 

We introduce the notation
\begin{equation}
  \label{eq:71}
  \sigmaj{j}
  =_{\bydefinition}
  k_1 \left(\uij{1}{j+1} - 2\uij{1}{j} +  \uij{1}{j-1} \right) - k_2 \left(\uij{1}{j} - \uij{2}{j}\right),
\end{equation}
for the total stress/force acting on the $j$-th unit cell, and using this notation we can rewrite~\eqref{eq:68} as
\begin{equation}
  \label{eq:72}
  \mij{1}{j} \ddd{\uij{1}{j}}{t}  - \sigmaj{j} = 0.
\end{equation}
Following the approach outlined in the previous section, we use the stress/force relation defined in~\eqref{eq:71}, and we get the following expression
\begin{equation}
  \label{eq:73}
  \uij{2}{j}
  =
  \frac{1}{k_2}
  \left(
    \sigmaj{j}
    -
    k_1 \left(\uij{1}{j+1} - 2\uij{1}{j} +  \uij{1}{j-1} \right)
  \right)
  +
  \uij{1}{j}
\end{equation}
for the inner mass displacement $\uij{2}{j}$ in terms of the stress/force $\sigmaj{j}$ and the outer mass displacements. We differentiate this equation twice with respect to time, and we get
\begin{equation}
  \label{eq:74}
  \ddd{}{t}
  \uij{2}{j}
  =
  \frac{1}{k_2}
  \ddd{}{t}
  \left(
    \sigmaj{j}
    -
    k_1 \left(\uij{1}{j+1} - 2\uij{1}{j} +  \uij{1}{j-1} \right)
  \right)
  +
  \ddd{}{t}
  \uij{1}{j}.
\end{equation}
On the left-hand side we use the evolution equation for the inner mass~\eqref{eq:69}, while on the right-hand side we make use of evolution equation for the outer mass rewritten in terms of the stress/force~\eqref{eq:72}, and we arrive at
\begin{equation}
  \label{eq:75}
  k_2 \left(\uij{1}{j} - \uij{2}{j}\right)
  =
  \frac{\mij{2}{j}}{k_2}
  \ddd{}{t}
  \left(
    \sigmaj{j}
    -
    k_1 \left(\uij{1}{j+1} - 2\uij{1}{j} +  \uij{1}{j-1} \right)
  \right)
  +
  \frac{\mij{2}{j}}{\mij{1}{j}}
  \sigmaj{j}.
\end{equation}
We note that in this setting, we have exploited the fact that the right-hand side in~\eqref{eq:68} is equal to zero, meaning that the system is not subject to ``volumetric forces''. 
Finally, we use~\eqref{eq:71}, and rewrite the left-hand side of the previous equation as
\begin{equation}
  \label{eq:76}
  -
  \left(
    \sigmaj{j}
    -
    k_1 \left(\uij{1}{j+1} - 2\uij{1}{j} +  \uij{1}{j-1} \right)
  \right)
  =
  \frac{\mij{2}{j}}{k_2}
  \ddd{}{t}
  \left(
    \sigmaj{j}
    -
    k_1 \left(\uij{1}{j+1} - 2\uij{1}{j} +  \uij{1}{j-1} \right)
  \right)
  +
  \frac{\mij{2}{j}}{\mij{1}{j}}
  \sigmaj{j}
  .
\end{equation}
This equation contains only the displacements of the outer masses $\mij{1}{j}$ and the stresses/forces~$\sigmaj{j}$.

In summary, we have shown that the system
\begin{subequations}
  \label{eq:77}
  \begin{align}
    \label{eq:78}
    \mij{1}{j} \ddd{\uij{1}{j}}{t}  - \sigmaj{j} &= 0, \\
    \label{eq:79}
    \frac{\mij{2}{j}}{k_2}
    \ddd{\sigmaj{j}}{t}
    +
    \left(
    1
    +
    \frac{\mij{2}{j}}{\mij{1}{j}}
    \right)
    \sigmaj{j}
                                                 &=
                                                   \frac{\mij{2}{j} k_1}{k_2}
                                                   \ddd{}{t}
                                                   \left(\uij{1}{j+1} - 2\uij{1}{j} +  \uij{1}{j-1} \right)
                                                   +
                                                   k_1 \left(\uij{1}{j+1} - 2\uij{1}{j} +  \uij{1}{j-1} \right),
  \end{align}
\end{subequations}
for unknowns $\uij{1}{j}$ and $\sigmaj{j}$ is equivalent to the system
\begin{subequations}
  \label{eq:80}
  \begin{align}
    \label{eq:81}
    \mij{1}{j} \ddd{\uij{1}{j}}{t}  - k_1 \left(\uij{1}{j+1} - 2\uij{1}{j} +  \uij{1}{j-1} \right) + k_2 \left(\uij{1}{j} - \uij{2}{j}\right)  &= 0, \\
    \label{eq:82}
    \mij{2}{j} \ddd{\uij{2}{j}}{t}  + k_2 \left(\uij{2}{j} -  \uij{1}{j}\right) & = 0,
  \end{align}
\end{subequations}
for unknowns $\uij{1}{j}$ and $\uij{2}{j}$. The reduced system~\eqref{eq:77} incorporates the effects due to the inner mass through the rate-type \emph{effective constitutive relation} for the stress/force and the strain/displacement, while the original system completely resolves the motion of the inner mass $\mij{2}{j}$ and uses simple linear constitutive relations for the individual springs. This simple calculation shows that the original model~\eqref{eq:80} can be rewritten in terms of the standard Newtonian balance of linear momentum for the outer mass with a novel effective constitutive relation. There is no need to introduce the notion of \emph{negative mass} nor to modify Newton's second law.

As expected, if we set the inner masses $\mij{2}{j}$ equal to zero, both models reduce to the standard one-dimensional lattice models, see, for example, \cite{brillouin.l:wave*1} or~\cite{kunin.ia:elastic*1,kunin.ia:elastic}. Furthermore, we can again identify the energy behind both models. The elastic stored energy for the $j$-th cell in the mass-in-mass one-dimensional lattice model is given by the formula
\begin{subequations}
  \label{eq:83}
  \begin{equation}
    \label{eq:84}
    \Wj{j} = \frac{1}{2} k_1 \left(\uij{1}{j+1} - \uij{1}{j} \right)^2 + \frac{1}{2} k_2 \left(\uij{2}{j} - \uij{1}{j} \right)^2,
  \end{equation}
  while the kinetic energy of the $j$-th cell reads
  \begin{equation}
    \label{eq:85}
    \Tj{j} = \frac{1}{2} \mij{1}{j} \left( \dd{\uij{1}{j}}{t} \right)^2 +  \frac{1}{2} \mij{2}{j} \left( \dd{\uij{2}{j}}{t} \right)^2,
  \end{equation}
  see~\cite{huang.hh.sun.ct.ea:on}.
\end{subequations}
Both these quantities can we rewritten in terms of the displacements of the outer masses $\mij{1}{l}$ and the stresses/forces~$\sigmaj{l}$, indeed, by virtue of~\eqref{eq:71} we get counterparts of~\eqref{eq:83} as
\begin{subequations}
  \label{eq:86}
  \begin{align}
    \label{eq:87}
    \Wj{j} &= \frac{1}{2} k_1 \left(\uij{1}{j+1} - \uij{1}{j} \right)^2
             +
             \frac{1}{2} k_2
             \left(
             \frac{
             \sigmaj{j}
             -
             k_1 \left(\uij{1}{j+1} - 2\uij{1}{j} +  \uij{1}{j-1} \right)
             }
             {
             k_2
             }
             \right)^2, \\
    \label{eq:88}
    \Tj{j} &= \frac{1}{2} \mij{1}{j} \left( \dd{\uij{1}{j}}{t} \right)^2
             +
             \frac{1}{2} \mij{2}{j}
             \left(
             \frac{1}{k_2}
             \left(
             \dd{
             \sigmaj{j}
             }{t}
             -
             k_1
             \dd{}{t}
             \left(\uij{1}{j+1} - 2\uij{1}{j} +  \uij{1}{j-1} \right)
             \right)
             +
             \dd{\uij{1}{j}}{t}
             \right)^2.
  \end{align}
\end{subequations}

\section{Dispersion relation for mass-in-mass lattice system}
\label{sec:disp-relat-mass}

The appealing feature of metamaterials is the \emph{band gap} phenomenon observed in wave transmission. Concerning the wave transmission in mass-in-mass one-dimensional lattice system~\eqref{eq:80} one can go back to the standard analysis in~\cite{huang.hh.sun.ct.ea:on} and identify the corresponding \emph{dispersion relation}.

\subsection{Standard approach}
\label{sec:standard-approach-1}
Following \cite{huang.hh.sun.ct.ea:on} we take the harmonic wave \emph{ansatz} in the form
\begin{equation}
  \label{eq:89}
  \uij{\gamma}{j+n} = B_\gamma \exponential{\iunit \left(qx + nqL - \omega t\right)},
\end{equation}
where $B_\gamma$ is the wave amplitude, $q$ is the wavenumber, $\omega$ is the angular frequency and $\gamma \in \left\{1, 2\right\}$, $n \in \Z$. Substituting the \emph{ansatz}~\eqref{eq:89} into the governing equations~\eqref{eq:80} for the $j$-th elementary cell we arrive at the system of algebraic equations 
\begin{subequations}
  \label{eq:90}
  \begin{align}
    \label{eq:91}
    - \omega^2 \mij{1}{j} B_1 - k_1 \left( \exponential{\iunit qL} - 2 + \exponential{-\iunit qL} \right) B_1 + k_2 \left(B_1 - B_2\right) &= 0, \\
    \label{eq:92}
    - \omega^2 \mij{2}{j} B_2 +  k_2 \left(B_2 - B_1\right) &= 0.
  \end{align}
\end{subequations}
Using the identity $ \exponential{\iunit qL} - 2 + \exponential{-\iunit qL} = 2 \left( \cos \left(qL\right) - 1 \right)$, we see that this system of equations can be rewritten in the form
\begin{equation}
  \label{eq:93}
  \begin{bmatrix}
    - \omega^2 \mij{1}{j} + 2k_1 \left( 1 - \cos \left(qL\right) \right) + k_2 & -k_2 \\
    -k_2 & -\omega^2\mij{2}{j}  + k_2
  \end{bmatrix}
  \begin{bmatrix}
    B_1 \\
    B_2
  \end{bmatrix}
  =
  \begin{bmatrix}
    0 \\
    0
  \end{bmatrix},
\end{equation}
hence the solvability condition for this system of equations reads
\begin{equation}
  \label{eq:94}
  \mij{1}{j}\mij{2}{j} \omega^4 - \left[ k_2\left( \mij{1}{j} + \mij{2}{j} \right) + 2k_1\mij{2}{j} \left( 1 - \cos \left(qL\right) \right) \right] \omega^2 + 2 k_1 k_2 \left( 1 - \cos \left(qL\right) \right) = 0.
\end{equation}
If the inner/outer masses in the elementary cells are identical, that is if $\mij{1}{j} =_{\bydefinition} m_1$ and  $\mij{2}{j} =_{\bydefinition} m_2$ for all $j \in \Z$, which is the case we focus on in the remaining part of the present study, then the single solvability condition~\eqref{eq:94} yields the sought dispersion relation in the form of single algebraic relation
\begin{gather}
  m_1 m_2 \omega^4 - \left[ k_2\left( m_1 + m_2 \right) + 2k_1 m_2 \left( 1 - \cos \left(qL\right) \right) \right] \omega^2 + 2 k_1 k_2 \left( 1 - \cos \left(qL\right) \right) = 0.   \label{eq:95}
\end{gather}
This is the dispersion relation derived by~\cite{huang.hh.sun.ct.ea:on}.

In the absence of the inner mass, $m_2 = 0$, the dispersion relation \eqref{eq:95} reduces to the standard dispersion relation for the monoatomic one-dimensional lattice system
\begin{equation}
  \label{eq:96}
  \omega^2 = \frac{2 k_1 \left( 1 - \cos \left(qL\right) \right)}{m_1}.
\end{equation}
Being inspired by the monoatomic dispersion relation~\eqref{eq:96}, one can try to convert the dispersion relation~\eqref{eq:95} into the form
\begin{equation}
  \label{eq:97}
  \omega^2 = \frac{2 k_1 \left( 1 - \cos \left(qL\right) \right)}{\meff}.
\end{equation}
This can be done provided that we set
\begin{equation}
  \label{eq:98}
  \meff
  =
  \msum
  +
  \frac{m_2 \left(\frac{\omega}{\omega_2}\right)^2}{1 - \left(\frac{\omega}{\omega_2}\right)^2},
\end{equation}
where $\msum =_{\bydefinition} m_1 + m_2$. As in the single mass-in-mass system, the effective mass is \emph{frequency dependent}, and for some values of~$\omega$, $\meff$ is even negative. One can thus claim, that the system described \emph{in the Fourier space} by the algebraic equations~\eqref{eq:93} with $\mij{1}{j} =_{\bydefinition} m_1$ and  $\mij{2}{j} =_{\bydefinition} m_2$ is---from the perspective of dispersion relation---equivalent to the monoatomic one-dimensional lattice system described by the dispersion relation
\begin{equation}
  \label{eq:99}
  - \omega^2 \meff + 2k_1 \left( 1 - \cos \left(qL\right) \right) = 0,
\end{equation}
provided that the effective mass $\meff$ is given by the formula~\eqref{eq:98}.

The mathematical calculations have been carried out correctly, however the interpretation is untenable from the physical perspective. We have to deal with frequency dependent and possibly negative effective mass. Furthermore, as in the case of simple mass-in-mass system, we also observe that due to the frequency dependence of the effective mass $\meff$, we \emph{can not claim} that~\eqref{eq:99} is a frequency space (Fourier space) counterpart of the governing equation for the monoatomic one-dimensional lattice system
\begin{equation}
  \label{eq:100}
   \meff \ddd{\uij{1}{j}}{t}  - k_1 \left(\uij{1}{j+1} - 2\uij{1}{j} +  \uij{1}{j-1} \right) = 0.
\end{equation}
Such an unsound physical interpretation (frequency dependent mass, negative mass, restriction to the frequency space) of the \emph{correct} mathematical manipulation however calls for a more appropriate and logical interpretation of the problem.

\subsection{Approach based on a rate-type constitutive relation}
\label{sec:approach-based-rate-1}

As in the case of the single mass-in-mass system, we must shift our attention to the force/stress versus displacement/strain constitutive relation. This is the same modus operandi as in going from~\eqref{eq:25} to~\eqref{eq:28} in the case of the single mass-in-mass system. We have shown that the governing equations for the mass-in-mass lattice system~\eqref{eq:80} can be rewritten in the form~\eqref{eq:77}. While the original system ``fully resolves'' the motion of the inner mass~$\mij{2}{j}$, the idea behind~\eqref{eq:77} is to interpret the dynamics due to the inner mass motion as a non-standard force/stress versus displacement/strain constitutive relation.

Since these systems are equivalent, they must give the same dispersion relation. For the sake of completeness we show this by an explicit calculation. We start with the harmonic wave \emph{ansatz}
\begin{subequations}
  \label{eq:101}
  \begin{align}
    \label{eq:102}
    \uij{1}{j+n} &= B_1 \exponential{\iunit \left(qx + nqL - \omega t\right)}, \\
    \label{eq:103}
    \sigmaj{j+n} &= S \exponential{\iunit \left(qx + nqL - \omega t\right)}.
  \end{align}
\end{subequations}
The notation is the same as in~\eqref{eq:89}. Substituting the \emph{ansatz} \eqref{eq:102}-\eqref{eq:103} into the governing equations~\eqref{eq:77} for the $j$-the elementary cell we arrive at the system of algebraic equations
\begin{subequations}
  \label{eq:104}
  \begin{align}
    \label{eq:105}
    - \omega^2 \mij{1}{j} B_1  - S &= 0, \\
    \label{eq:106}
    - \omega^2
    \frac{\mij{2}{j}}{k_2}
    S
    +
    \left(
    1
    +
    \frac{\mij{2}{j}}{\mij{1}{j}}
    \right)
    S
                                   &=
                                     2
                                     \frac{\mij{2}{j} k_1}{k_2}
                                     \omega^2
                                     \left(1 - \cos \left(qL\right)\right)
                                     B_1
                                     -
                                     2k_1 \left( 1 - \cos \left(qL\right) \right) B_1.
  \end{align}
\end{subequations}
This system can be rewritten in the matrix form
\begin{equation}
  \label{eq:107}
  \begin{bmatrix}
    - \omega^2 \mij{1}{j}
    &
      -1
    \\
    -2 \left( \frac{\mij{2}{j} k_1}{k_2} \omega^2 - k_1 \right)
    \left(1 - \cos \left(qL\right)\right)
    &
      - \omega^2 \frac{\mij{2}{j}}{k_2}
      +
      \left(
      1
      +
      \frac{\mij{2}{j}}{\mij{1}{j}}
      \right)
  \end{bmatrix}
  \begin{bmatrix}
    B_1 \\
    S
  \end{bmatrix}
  =
  \begin{bmatrix}
    0 \\
    0
  \end{bmatrix}
  ,
\end{equation}
and the solvability condition for this system of equations reads
\begin{equation}
  \label{eq:108}
  \mij{1}{j} \mij{2}{j} \omega^4 
  -
  \left[
    k_2
    \left(\mij{1}{j} + \mij{2}{j}\right)
    +
    2 k_1 \mij{2}{j} 
    \left(1 - \cos \left(qL\right)\right)
  \right]
  \omega^2
  +
  2 k_1 k_2
  \left(1 - \cos \left(qL\right)\right)
  =
  0.
\end{equation}
This is the same equation as~\eqref{eq:94}. If the inner/outer masses in the elementary cells are identical, that is if $\mij{1}{j} =_{\bydefinition} m_1$ and  $\mij{2}{j} =_{\bydefinition} m_2$ for all $j \in \Z$, which is the case we are interested in, then the single solvability condition~\eqref{eq:108} yields the sought dispersion relation in the form of single algebraic relation
\begin{equation}
  m_1 m_2 \omega^4 - \left[ k_2\left( m_1 + m_2 \right) + 2k_1 m_2 \left( 1 - \cos \left(qL\right) \right) \right] \omega^2 + 2 k_1 k_2 \left( 1 - \cos \left(qL\right) \right) = 0.
\end{equation}
The comparison with~\eqref{eq:95} reveals that this is indeed the same dispersion relation arrived at appealing to the physically erroneous concept of negative mass.

\section{Continuous model}
\label{sec:cont-repr-1}
The discrete one-dimensional mass-in-mass lattice system can be seen as a precursor for a one-dimensional homogeneous continuous solid model. The standard naive identification of a corresponding spatially homogeneous continuous model is based on the approximation
\begin{equation}
  \label{eq:109}
  \ppd{u}{x} \approx \frac{\uij{1}{j+1} - 2\uij{1}{j} +  \uij{1}{j-1}}{L^2},
\end{equation}
see~\cite{brillouin.l:wave*1}, \cite{kunin.ia:elastic*1,kunin.ia:elastic}, \cite{berezovski.a:elastic} and a historical account in~\cite{challamel.n.zhang.yp.ea:discrete}. We use the term ``naive identification'' here because the ``equivalent'' continuous model is known to have a different dispersion relation than the continuous one, see~\cite{challamel.n.zhang.yp.ea:discrete} for remarks and \cite{filimonov.am.kurchanov.pf.ea:some}, \cite{filimonov.am:some,filimonov.am:continuous}, \cite{andrianov.iv:specific} and~\cite{challamel.n.picandet.v.ea:revisiting} for an in-depth discussion. 

The naive identification \eqref{eq:109} is followed in~\cite{huang.hh.sun.ct.ea:on}, and if we do the same, applying~\eqref{eq:109} to the system~\eqref{eq:78} with identical masses $\mij{1}{j} =_{\bydefinition} m_1$ and  $\mij{2}{j} =_{\bydefinition} m_2$ for all $j \in \Z$, we obtain
\begin{subequations}
  \label{eq:110}
  \begin{align}
    \label{eq:111}
    m_1 \ppd{u}{t}  - \sigma &= 0, \\
    \label{eq:112}
    \frac{m_2}{k_2}
    \ppd{\sigma}{t}
    +
    \left(
    1
    +
    \frac{m_2}{m_1}
    \right)
    \sigma
                                                 &=
                                                   \frac{m_2 k_1}{k_2} 
                                                   \ppd{}{t}
                                                   \left(
                                                   L^2
                                                   \ppd{u}{x}
                                                   \right)
                                                   +
                                                   k_1 L^2\ppd{u}{x}.
  \end{align}
\end{subequations}
The initial conditions are imposed for the displacement $u$ as in the standard monoatomic lattice model, while for $\sigma$ we have to prescribe the initial conditions for $\sigma$ and the time derivative $\pd{\sigma}{t}$, see the similar discussion in Section~\ref{sec:approach-based-rate}. 

We can use \eqref{eq:86} to identify the stored energy, 
\begin{equation}
  \label{eq:stored}
  W =_{\bydefinition} \frac{1}{2} k_1 L^2 \left ( \pd{u}{x} \right )^2 + \frac{1}{2k_2} \left ( \sigma - k_1 L^2 \ppd{u}{x} \right )^2, 
\end{equation} 
and the kinetic energy
\begin{equation}
   \label{eq:kinetic}
   T =_{\bydefinition} \frac{1}{2}m_1 \left ( \pd{u}{t} \right )^2 + \frac{1}{2}m_2
   \left[
     \frac{1}{k_2} \left ( \pd{\sigma}{t} - k_1 L^2 \pd{}{t} \ppd{u}{x} \right ) + \pd{u}{t}
   \right]^2, 
\end{equation}
associated to the continuous system~\eqref{eq:110}. In particular, we compute  
\begin{equation}
  \label{eq:dtW}
  \pd{W}{t}
  =
  k_1 L^2 \pd{u}{x} \pd{^2 u}{t \partial x}
  + 
  \left (\sigma - k_1 L^2 \ppd{u}{x} \right )
  \frac{1}{k_2}
  \left (
    \pd{\sigma}{t} - k_1 L^2 \pd{}{t} \ppd{u}{x} 
  \right ), 
\end{equation}
and using \eqref{eq:110}, we compute 
\begin{multline}
  \label{eq:dtT}
  \pd{T}{t}
  =
  m_1 \ppd{u}{t} \pd{u}{t}
  + 
  \left[ 
    \frac{m_2}{k_2}
    \left (
      \ppd{\sigma}{t} - k_1 L^2 \ppd{}{t} \ppd{u}{x}
    \right )
    +
    m_2 \ppd{u}{t}
  \right]
  \left[
    \frac{1}{k_2} \left ( \pd{\sigma}{t} - k_1 L^2 \pd{}{t} \ppd{u}{x} \right ) + \pd{u}{t}
  \right]
  \\
  =
  \sigma \pd{u}{t}
  +
  \left[
    -\sigma + k_1 L^2 \ppd{u}{x}
  \right]
  \left[
    \frac{1}{k_2} \left ( \pd{\sigma}{t} - k_1 L^2 \pd{}{t} \ppd{u}{x} \right ) + \pd{u}{t}
  \right]
  \\
  =
  k_1 L^2 \ppd{u}{x} \pd{u}{t}
  -
  \left(
    \sigma - k_1 L^2 \ppd{u}{x}
  \right)
  \frac{1}{k_2}
  \left(
    \pd{\sigma}{t} - k_1 L^2 \pd{}{t} \ppd{u}{x} 
  \right). 
\end{multline}
Adding \eqref{eq:dtW} to \eqref{eq:dtT}, integrating over $[a,b]$ with respect to $x$, and integrating by parts, we obtain a statement of \textit{balance of the total energy} for the system \eqref{eq:110} and the material segment $[a,b]$, 
\begin{equation}
	\dd{}{t} \int_{x=a}^b \left(T + W\right) \, \diff x =  k_1 L^2 \left. \pd{u}{x} \pd{u}{t} \right |_{x = a}^{x = b}. \label{eq:balanceenergy}
\end{equation}
The term on the right-hand side of \eqref{eq:balanceenergy} is the power expended by the segments $(-\infty,a]$ and $[b,\infty)$ on $[a,b]$, and there are no additional terms expressing dissipation of energy. In particular, taking the limit $a \rightarrow -\infty$ and $b \rightarrow \infty$, and assuming appropriate decay of $u$ as $|x| \rightarrow \infty$, we obtain \emph{conservation of energy},
\begin{equation}
	\dd{}{t} \int_{x=-\infty}^\infty \left(T + W\right) \, \diff x = 0. \label{eq:conservationeenergy}
\end{equation}
 
\section{Conclusion}
\label{sec:conclusion}
We have used the concept of an \emph{effective constitutive relation} to interpret the governing equations for a metamaterial without frequency dependent and possibly negative \emph{effective mass}. The effective constitutive relation has been identified as a rate-type constitutive relation relating the force/stress and the displacement/strain. We reiterate that the concept of rate-type constitutive relations is well established in the theory of viscoelastic response, wherein a complex arrangement of springs and dashpots is described by an effective rate-type constitutive relation
\begin{equation}
  \label{eq:113}
  f \left( \varepsilon, \sigma, \dd{\varepsilon}{t}, \dd{\sigma}{t}, \ddd{\varepsilon}{t}, \ddd{\sigma}{t}, \cdots \right) = 0.
\end{equation}
However, it is imperative to recognize that the rate-type constitutive relations that we have discussed characterizes an \emph{elastic} solid, not a \emph{viscoelastic} solid. Also note that~\eqref{eq:113} is in general an implicit constitutive relation and it is a generalization of the algebraic implicit constitutive relation for elastic bodies, see~\cite{rajagopal.kr:on*3}. In the same context, see also the concept of \emph{material with fading memory}, which has been gainfully employed to develop constitutive relations in non-Newtonian fluid mechanics, see~\cite{coleman.bd.noll.w:approximation} and~\cite{prusa.rajagopal.kr:on}. These rate-type constitutive relations can be also used in the modeling of inelastic phenomena such as plasticity or the Mullins effect, see, for example, \cite{rajagopal.kr.srinivasa.ar:inelastic,rajagopal.kr.srinivasa.ar:implicit*1}. Moreover, the naive one-dimensional models based on the rate-type equation can be recast in the fully three-dimensional thermodynamically consistent setting, see, for example, \cite{malek.j.rajagopal.kr.ea:on}, \cite{cichra.d.prusa.v:thermodynamic}, \cite{cichra.d.gazca-orozco.pa.ea:thermodynamic}.

The rate-type constitutive relations~\eqref{eq:44}, \eqref{eq:79}, and~\eqref{eq:112} identified in this work bring an interesting new theme into the well-established framework of rate-type constitutive relations. They give a clue how to design a non-trivial effective rate-type constitutive relation that is \emph{conservative}, that is, possessing an associated conserved energy depending on the force/stress and its time derivatives, see \eqref{eq:conservationeenergy}. In short, the rate-type constitutive relations obtained in this study might be seen as a revival of \emph{hypoelastic materials} in the nomenclature by~\cite{truesdell.c:hypo-elasticity} and \cite{truesdell.c.noll.w:non-linear}. (See also~\cite{noll.w:on}, \cite{bernstein.b.ericksen.jl:work}, \cite{bernstein.b:relations,bernstein.b:hypo-elasticity}, \cite{xiao.h.bruhns.ot.ea:hypo-elasticity}, \cite{leonov.ai:on*1} and \cite{bernstein.b.rajagopal.k:thermodynamics} to name a few developments in the field.) The three dimensional nonlinear models for finite deformations that would correspond to the one-dimensional models studied in this work are however yet to be investigated.


\bibliographystyle{chicago}
\bibliography{vit-prusa}

\end{document}